\documentclass[journal]{IEEEtran}
\usepackage{multirow}
\usepackage{diagbox}
\usepackage{amssymb}
\usepackage{amsmath}
\usepackage{graphicx}
\usepackage{cite}
\usepackage{citesort}
\usepackage{subfigure}
\usepackage{graphicx,epstopdf}
\usepackage{epsfig}	
\usepackage{cite,graphicx,amsmath,amssymb}
\usepackage{comment}

\usepackage{amssymb}
\usepackage{amsmath}
\usepackage{cite}
\usepackage{url}
\usepackage{xcolor}
\usepackage{cite,graphicx,amsmath,amssymb}
\usepackage{subfigure}
\usepackage{citesort}
\usepackage{fancyhdr}
\usepackage{mdwmath}
\usepackage{mdwtab}
\usepackage{caption}
\usepackage{amsthm}
\usepackage{lipsum}

\newtheorem{theorem}{Theorem}

\newtheorem{lemma}{Lemma}

\newtheorem{corollary}{Corollary}

\newtheorem{remark}{Remark}  

\makeatletter
\def\ScaleIfNeeded{%
\ifdim\Gin@nat@width>\linewidth \linewidth \else \Gin@nat@width
\fi } \makeatother

\begin{document}

\title{Outage Behaviors of NOMA-based Satellite Network over Shadowed-Rician Fading Channels}

\author{Xinwei~Yue,~\IEEEmembership{Member,~IEEE,} Yuanwei\ Liu,~\IEEEmembership{Senior~IEEE,} Yuanyuan~Yao,~\IEEEmembership{Member,~IEEE,} Tian~Li,~\IEEEmembership{Member,~IEEE,} Xuehua Li,~\IEEEmembership{Member,~IEEE,} Rongke Liu,~\IEEEmembership{Senior~IEEE,} and Arumugam~Nallanathan,~\IEEEmembership{Fellow,~IEEE}

\thanks{X. Yue, Y. Yao and X. Li are with the Key Laboratory of Modern Measurement $\&$ Control Technology, Ministry of Education and also with the School of Information and Communication Engineering, Beijing Information Science and Technology University, Beijing 100101, China. (email: \{xinwei.yue, yyyao, lixuehua\}@bistu.edu.cn).}
\thanks{Y. Liu and A. Nallanathan are with the School of Electronic Engineering and Computer Science, Queen Mary University of London, London E1 4NS, U.K. (email: \{yuanwei.liu, a.nallanathan\}@qmul.ac.uk).}
\thanks{Tian Li is with the 54th Research Institute of China Electronics Technology Group Corporation, Shijiazhuang Hebei 050081, China and also with Beijing University of Posts and Telecommunications, Beijing 100876, China (email: t.li@ieee.org).}
\thanks{R. Liu is with the School of Electronic and Information Engineering, Beihang University, Beijing 100191, China (email: rongke$\_$liu@buaa.edu.cn).}
}

\maketitle

\begin{abstract}
This paper investigates the application of non-orthogonal multiple access (NOMA) to satellite communication network over Shadowed-Rician fading channels. The impact of imperfect successive interference cancellation (ipSIC) on NOMA-based satellite network is taken into consideration from the perspective of practical scenarios.  We first derive new exact expressions of outage probability for the $p$-th terrestrial user and provide the corresponding asymptotic analysis results. The diversity order of $zero$ and $p$ are achieved by the $p$-th terrestrial user with ipSIC and perfect successive interference cancellation (pSIC), respectively.
Finally, the presented simulation results show that: 1) On the condition of pSIC, the outage behaviors of NOMA-based satellite network are superior to that of orthogonal multiple access; 2) With the value of residual interference increasing, the outage performance of terrestrial users with ipSIC is becoming worse seriously; and 3) Infrequent light shadowing of Shadowed-Rician fading brings the better outage probability compared to frequent heavy and average shadowing.
\end{abstract}


\section{Introduction}
With development of Internet of Things (IoT) and satellite communication business needs, the IoT-based satellite networks have received a vast amount of attention, which has been viewed as a crucial application scenarios. Currently, a large number of satellite machine-type-communication terminals on the ground have introduced a great challenge to multiple access for the satellite communication networks. Non-orthogonal multiple access (NOMA) is an effective approach to meet the requirements of massive connections \cite{Liu8114722Beyond}. 

Until now, the use of NOMA has been confirmed to have better performance gain from the perspective of improving the spectral efficiency and user fairness \cite{Ding2014Randomly,Choi6708131}. By extending the concept of NOMA to cooperative communication, the authors of \cite{Ding2014Cooperative,Liu7445146SWIPT,Yue8026173FD} discussed the users' outage behaviors by regarding the nearby user as a relay. To present the valuable insights of security performance, the authors in \cite{Liu7812773Security,Zhao8PLSNOMA,Lei2018SecuringTrusted} have investigated the secrecy outage probability and maximize the sum secrecy rate of NOMA systems by invoking joint precoding optimization. Furthermore, a pair of NOMA assisted caching strategies were proposed to emphasize the wireless caching \cite{Ding8368286Caching}. To better understand NOMA assisted unmaned aerial vehicle networks, the authors of \cite{Liu8641425UAV,Zhao8UAVNOMA} have evaluated the performance in terms of both outage probability and fly trajectory.

Satellite networks are expected to be a complementary role of terrestrial communication systems, since it is capable of supplying the spacious coverage and short deployment time \cite{Cioni8473482Satellite,Zhang8700136}. Hence applying NOMA technology to satellite communications will be further a promising way to extend the applications of the integration between space and earth. To evaluate the performance of NOMA satellite networks, the authors in \cite{Yan8374960Land} researched the outage probability of ground users, while it did not consider the outage probability performance of any one of $M$ users under the condition of order statistics. From the view of practical scenarios, the SIC procedure exists the potential concrete issues i.e., error propagation and complexity scaling, which will result in errors in decoding process. Hence it is important to take these undesirable influence from imperfect successive interference cancellation (ipSIC) into consideration \cite{Yue8370069Unified}. To the best of our knowledge, the outage behaviors of terrestrial users with ipSIC have not been well evaluated.

Triggered by these treatises, we derive new exact and asymptotic expressions of outage probability for the ordered terrestrial users. The impact of channel parameters on NOMA-based satellite network is discussed in detail. Numerical results corroborate our analyses that: 1) On the condition of pSIC, the outage behaviors of NOMA-based satellite network are superior to that of orthogonal multiple access (OMA); 2) With the increasing of residual interference, the outage behaviors of terrestrial users with ipSIC are becoming worse seriously; and 3) Infrequent light shadowing of Shadowed-Rician fading results in a reduced outage probability.
\section{Network Model}\label{Network Model}
Consider a NOMA-based satellite communication scenario, where a satellite broadcasts the superposed information to terrestrial users. Assuming that $M$ users randomly distribute within the coverage of the satellite. The satellite and terrestrial users are equipped single antenna, respectively. The Shadowed-Rician model is employed to describe satellite land links. We assume that ${h_p}$ denotes the channel fading coefficient from satellite to the $p$-th terrestrial user.
Without loss of generality, the corresponding channel gains between the satellite and $M$ terrestrial users are ordered as ${\left| {{h_1}} \right|^2} \le {\left| {{h_2}} \right|^2} \le  \cdots  \le {\left| {{h_{M - 1}}} \right|^2} \le {\left| {{h_M}} \right|^2}$ \cite{David2003Order}. These satellite-users links are disturbed by additive white Gaussian noise (AWGN) with mean power ${\tilde N}$.
In light of the above assumptions, the cumulative distribution function (CDF) and probability density function of channel gains from satellite to the $p$-th user under the unordered conditions are given by \cite{Abdi1198102}
\begin{align}\label{the CDF of unsorted channel gains}
{F_{{{\left| {{{\hat h}_p}} \right|}^2}}}\left( x \right) = {\alpha _p}\sum\limits_{k = 0}^\infty  {\frac{{{{\left( {{m_p}} \right)}_k}\delta _p^k}}{{{{\left( {k!} \right)}^2}\beta _p^{k + 1}}}} \gamma \left( {k + 1,x{\beta _p}} \right),
\end{align}
and
\begin{align}\label{the PDF of unsorted channel gains}
{f_{{{\left| {{{\hat h}_p}} \right|}^2}}}\left( x \right)= & \frac{1}{{2{b_p}}}{\left( {\frac{{2{b_p}{m_p}}}{{2{b_p}{m_p} + {\Omega _p}}}} \right)^{{m_p}}}{e^{ - \frac{x}{{2{b_p}}}}} \nonumber \\
 & \times {}_1{F_1}\left( {{m_p};1;\frac{{x{\Omega _p}}}{{2{b_p}\left( {2{b_p}{m_p} + {\Omega _p}} \right)}}} \right)  ,
\end{align}
respectively, where $\gamma \left( {a,x} \right) = \int_0^x {{t^{a - 1}}{e^{ - t}}dt} $ denotes the lower incomplete Gamma function \cite[Eq. (8.350.1)]{gradshteyn}. ${\left( m \right)_k} = \Gamma \left( {m + k} \right)/\Gamma \left( m \right)$ is the Pochhammer symbol and $\Gamma \left( m \right)$ denotes the Gamma function.
 ${\delta _p} = {{{{{\Omega _p}} \mathord{\left/
 {\vphantom {{{\Omega _p}} {2{b_p}}}} \right.
 \kern-\nulldelimiterspace} {2{b_p}}}} \mathord{\left/
 {\vphantom {{{{{\Omega _p}} \mathord{\left/
 {\vphantom {{{\Omega _p}} {2{b_p}}}} \right.
 \kern-\nulldelimiterspace} {2{b_p}}}} {\left( {2{b_p}{m_p} + {\Omega _p}} \right)}}} \right.
 \kern-\nulldelimiterspace} {\left( {2{b_p}{m_p} + {\Omega _p}} \right)}}$, ${\alpha _p} = {{{{\left( {{{2{b_p}{m_p}} \mathord{\left/
 {\vphantom {{2{b_p}{m_p}} {\left( {2{b_p}{m_p} + {\Omega _p}} \right)}}} \right.
 \kern-\nulldelimiterspace} {\left( {2{b_p}{m_p} + {\Omega _p}} \right)}}} \right)}^{{m_p}}}} \mathord{\left/
 {\vphantom {{{{\left( {{{2{b_p}{m_p}} \mathord{\left/
 {\vphantom {{2{b_p}{m_p}} {\left( {2{b_p}{m_p} + {\Omega _p}} \right)}}} \right.
 \kern-\nulldelimiterspace} {\left( {2{b_p}{m_p} + {\Omega _p}} \right)}}} \right)}^{{m_p}}}} {2{b_p}}}} \right.
 \kern-\nulldelimiterspace} {2{b_p}}}$, and ${\beta _p} = 1/2{b_p}$.
 $\Omega _{p}$ and ${2{b_{p}}}$ denotes the average power of the line of sight (LoS) component and multipath component, respectively.
${{m_{p}}}$ is the Nakagami-$m$ parameter ranging from zero to infinite. 
${}_1{F_1}\left( {a;b;x} \right)$ denotes the confluent hypergeometric function \cite[Eq. (9.100)]{gradshteyn}.

In NOMA-based satellite network, the satellite transmits the superposed signals to multiple terrestrial users. Hence the received signal ${y_p}$ of the $p$-th user can be written as
\begin{align}\label{The received signal expression}
{y_p} = {h_p}\sum\limits_{i = 1}^M {\sqrt {{\eta _i}{G_s}{G_i}\left( {{\varphi _i}} \right){a_i}{P_s}} {x_i} + {n_p}},
\end{align}
where the power allocation factor $a_i$ of the $i$-th user satisfies $\mathop \sum \limits_{i = 1}^M {a _i} = 1$ and ${a_1} \ge {a_2} \ge  \cdots \ge {a_{M - 1}} \ge {a_{M}}$ to ensure the users' fairness. It is worth noting that the optimal power sharing strategy is capable of enhancing the performance of the network, which will be taken into account in the future work. The normalized transmission power at the satellite is denoted by $P_{s}$. The $i$-th user's signal $x_{i}$ is assumed to have zero mean and unit variance and ${n_p} \sim {\cal C}{\cal N}( {0,{\tilde N}} )$ denotes AWGN. In addition, ${\eta _i} = {\left( {{\lambda  \mathord{\left/{\vphantom {\lambda  {4\pi {d_i}}}} \right.\kern-\nulldelimiterspace} {4\pi {d_i}}}} \right)^2}$ denotes the free space loss coefficient of one beam. The wavelength is $\lambda  = {C\mathord{\left/
{\vphantom {C {{f_c}}}} \right.\kern-\nulldelimiterspace} {{f_c}}}$, where $C$ and $f_c$ denote the light speed and frequency, respectively. $d_i$ denotes the distance between the satellite and the $i$-user. ${G_s}$ is the antenna gain at the satellite. Given the $i$-th user's location, $\varphi _i$ represents the angle between it and beam center compared to the satellite. Hence the beam gain ${G_i}\left( {{\varphi _i}} \right)$ is given by \cite{Zheng6184256}
\begin{align}\label{The antenna gain at i-th user}
{G_i}\left( {{\varphi _i}} \right) = {G_i}\left( {\frac{{{J_1}\left( {{u_i}} \right)}}{{2{u_i}}} + 36\frac{{{J_3}\left( {{u_i}} \right)}}{{u_i^3}}} \right),
\end{align}
where $G_i$ denotes the antenna gain of the $i$-th user and ${u_i} = 2.07123{{\sin {\varphi _j}} \mathord{\left/ {\vphantom {{\sin {\varphi _j}} {\sin {\varphi _{j3{\rm{dB}}}}}}} \right.\kern-\nulldelimiterspace} {\sin {\varphi _{j3{\rm{dB}}}}}}$. ${\varphi _{j3{\rm{dB}}}}$ is the constant 3-dB angle for the beam. ${J_1}\left(  \cdot  \right)$ and ${J_3}\left(  \cdot  \right)$ denote the first-kind Bessel function with order one and three, respectively.

Following NOMA procedures \cite{Ding2014Randomly}, the received signal to interference and noise ratio (SINR) at the $p$-th user to detect the information of the $q$-th user $( p > q)$ is given by
\begin{align}\label{The SINR expression of the p-th user to detect the q-th user}
{\gamma _{p \to q}} = \frac{{{\phi _p}\rho {{\left| {{h_p}} \right|}^2}{a_q}}}{{{\phi _p}\rho {{\left| {{h_p}} \right|}^2}\sum\limits_{i = q + 1}^M {{a_i}} { + \eta \rho {{\left| {{h_I}} \right|}^2}} + 1}},
\end{align}
where ${\rho}=\frac{{{P_s}}}{{\tilde N}}$ denotes the transmit SNR, ${\phi _p} = {\eta _p}{G_s}{G_p}\left( {{\varphi _p}} \right)$ and $\eta   \in \left[ {0,1} \right]$. $\eta =1$ and $\eta =0$ denote ipSIC and pSIC, respectively. Without loss of generality, we assume that ${h_I}$ denotes the residual interference, which follows a complex Gaussian distribution with zero mean and variance ${\Omega _I}$ i.e.,  ${{h_I}}  \sim {\cal C}{\cal N}\left( {0,{\Omega _I}} \right)$.

Then the received SINR of $p$-th user\footnote{It is worth noting that the first user (i.e., $p=1$) with the worse channel condition does not perform SIC operation. Hence there is no residual interference term $\eta \rho {{\left| {{h_I}} \right|}^2}$ in \eqref{The SINR expression of the p-th user}.} detect the information by treating $M-p$ users' signals as interference is given by
\begin{align}\label{The SINR expression of the p-th user}
{\gamma _p} = \frac{{{\phi _p}\rho {{\left| {{h_p}} \right|}^2}{a_p}}}{{{\phi _p}\rho {{\left| {{h_p}} \right|}^2}\sum\limits_{i = p + 1}^M {{a_i}}  + \eta \rho {{\left| {{h_I}} \right|}^2} + 1}}.
\end{align}

After the information of $M-1$ users can be detected, the received SINR for the $M$-th user is given by
\begin{align}\label{The SINR expression of the M-th user}
{\gamma _M} = \frac{{\rho {a_M}{{\left| {{h_M}} \right|}^2}{\phi _M}}}{{\eta \rho {{\left| {{h_I}} \right|}^2} + 1}}.
\end{align}
 
\section{performance evaluation}\label{Performance evaluation}
\subsubsection{Outage Probability}
The SIC is carried out at the $p$-th user by detecting and canceling the $i$-th user's information $\left( {i \le p} \right)$ before it decodes its own signal. If the $p$-th user cannot detect the $i$-th user’s information, outage occurs and is denoted by ${{\rm{E}}_{p,i}}$. Hence the outage probability of $p$-th user can be formulated as follows:
\begin{align}\label{Op expression with ipSIC}
P_{}^p =& 1 - {\rm{Pr}}\left[ {{\rm{E}}_{p,1}^c \cap {\rm{E}}_{p,2}^c \cap  \cdots  \cap {\rm{E}}_{p,p}^c} \right],
\end{align}
where ${\rm{E}}_{p,i}^c$ denotes the complement event of ${{\rm{E}}_{p,i}}$.

\begin{theorem}\label{Theorem:CD-NOMA:the OP of p-th user with ipSIC}
Under the condition of ipSIC scheme, the exact expression of outage probability for the $p$-th terrestrial user in NOMA-based satellite network is given by
\begin{align}\label{CD-NOMA:the OP of p-th user with ipSIC}
P_{ipSIC}^p = &\frac{{{\Theta _p}}}{{{\Omega _I}}}\sum\limits_{l = 0}^{M - p} {{
   {M - p}  \choose
   l  }} \frac{{{{\left( { - 1} \right)}^l}\alpha _p^{p + l}}}{{p + l}}\int_0^\infty  {\left[ {\sum\limits_{k = 0}^\infty  {\frac{{{{\left( {{m_p}} \right)}_k}}}{{{{\left( {k!} \right)}^2}}}} } \right.} \nonumber \\
& {\left. { \times \frac{{\delta _p^k}}{{\beta _p^{k + 1}}}\gamma \left( {k + 1,\psi _p^ * \left( {\eta \rho x + 1} \right){\beta _p}} \right)} \right]^{p + l}}{e^{ - \frac{x}{{{\Omega _I}}}}}dx,
\end{align}
where ${\Theta _p} = \frac{{M!}}{{\left( {M - p} \right)!\left( {p - 1} \right)!}}$, $\psi _p^ *  = \max \left\{ {{\psi _1},...,{\psi _p}} \right\}$, ${\psi _p} = \frac{{{\gamma _{t{h_p}}}}}{{\rho {\phi _p}\left( {{a_p} - {\gamma _{t{h_p}}}\sum\limits_{i = p + 1}^M {{a_i}} } \right)}}$ with ${a_p} > {\gamma _{t{h_p}}}\sum\limits_{i = p + 1}^M {{a_i}} $, ${\psi _M} = \frac{{{\gamma _{t{h_M}}}}}{{\rho {\phi _M}{a_M}}}$, ${\gamma _{t{h_p}}} = {2^{ {{\tilde R}_p}}} - 1$ with $ {{\tilde R}_p}$ being the target data rate at the $p$-th user to detect $x_p$.
\end{theorem}
\begin{proof}
On the basis of \eqref{The SINR expression of the p-th user to detect the q-th user}, \eqref{The SINR expression of the p-th user}, and \eqref{The SINR expression of the M-th user}, \eqref{Op expression with ipSIC} can be calculated as follows:
\begin{align}\label{The derived process with ipSIC}
&P_{ipSIC}^p = 1 - {\rm{Pr}}\left[ {{{\left| {{h_p}} \right|}^2} > \psi _p^ * \left( {\eta \rho {{\left| {{h_I}} \right|}^2} + 1} \right)} \right] \nonumber\\
&  = 1 - \int_0^\infty  {\int_{\psi _p^ * \left( {\varpi \rho x + 1} \right)}^\infty  {{f_{{{\left| {{h_I}} \right|}^2}}}\left( x \right){f_{{{\left| {{h_p}} \right|}^2}}}\left( y \right)dydx} }\nonumber\\
&   = \int_0^\infty  {{F_{{{\left| {{h_p}} \right|}^2}}}\left( {\psi _p^ * \left( {\eta \rho x + 1} \right)} \right)\frac{1}{{{\Omega _I}}}{e^{ - \frac{x}{{{\Omega _I}}}}}dx}.
\end{align}
Based on \cite{David2003Order}, the relationship of CDF between the ordered channel gain and unordered channel gain can be expressed as
\begin{align}\label{The relationship of CDF takecare}
 {F_{{{\left| {{h_p}} \right|}^2}}}\left( x \right) =& \frac{{M!}}{{\left( {p - 1} \right)!\left( {M - p} \right)!}}\mathop \sum \limits_{l = 0}^{M - p} {M-p \choose
  l   } \nonumber\\
  &\times \frac{{{{\left( { - 1} \right)}^l}}}{{p + l}}{\left( {F_{\left| {{{\hat h}_p}} \right|}}\left( x \right)\right)^{p + l}},
\end{align}
where ${F_{\left| {{{\hat h}_p}} \right|}}\left( x \right)$ is the CDF of unsorted channel gain. Upon substituting \eqref{the CDF of unsorted channel gains} into \eqref{The relationship of CDF takecare} and combining \eqref{The derived process with ipSIC}, we can obtain \eqref{CD-NOMA:the OP of p-th user with ipSIC}.  The proof is completed.
\end{proof}


\begin{corollary}\label{Theorem:CD-NOMA:the OP of far user}
Under the condition of pSIC scheme, the closed-form expression of outage probability for the $p$-th user in NOMA-based satellite network can be given by
\begin{align}\label{CD-NOMA:the OP of p-th user}
P_{pSIC}^p =& {\Theta _p}\sum\limits_{l = 0}^{M - p} {{
   {M - p}  \choose
   l  }} \frac{{{{\left( { - 1} \right)}^l}\alpha _p^{p + l}}}{{p + l}} \nonumber \\
  &\times {\left[ {\sum\limits_{k = 0}^\infty  {\frac{{{{\left( {{m_p}} \right)}_k}\delta _p^k}}{{{{\left( {k!} \right)}^2}\beta _p^{k + 1}}}} \gamma \left( {k + 1,\psi _p^ * {\beta _p}} \right)} \right]^{p + l}} .
\end{align}
\end{corollary}

\subsubsection{Diversity order}
To get more insights, the diversity order is usually selected to be a metric, which is capable of describing how fast outage probability decreases with the transmit SNRs \cite{Ding2014Randomly,laneman2004cooperative}. Based on these explanations, the diversity order of terrestrial user can be given by
\begin{align}\label{The definiation of diversity order}
d =  - \mathop {\lim }\limits_{\rho  \to \infty } \frac{{\log \left( {P^\infty \left( \rho  \right)} \right)}}{{\log \rho }},
\end{align}
where ${P^\infty \left( \rho  \right)}$ denotes the asymptotic outage probability.

\begin{corollary}\label{Corollary:the expression of asymptotic OP for the p-th user}
The asymptotic outage probability of the $p$-th terrestrial user with ipSIC in the high SNR regime is given by
\begin{align}\label{The asymptotic expression of the p-th user with ipSIC}
P_{ipSIC}^{p,\infty } = &\frac{{{\Theta _p}}}{{{\Omega _I}}}\sum\limits_{l = 0}^{M - p} {{
   {M - p}  \choose
   l  }} \frac{{{{\left( { - 1} \right)}^l}\alpha _p^{p + l}}}{{p + l}}\int_0^\infty  {\left[ {\sum\limits_{k = 0}^\infty  {\frac{{{{\left( {{m_p}} \right)}_k}}}{{{{\left( {k!} \right)}^2}}}} } \right.} \nonumber \\
 &{\left. { \times \frac{{\delta _p^k}}{{\beta _p^{k + 1}}}\gamma \left( {k + 1,\eta x\vartheta _p^ * {\beta _p}} \right)} \right]^{p + l}}{e^{ - \frac{x}{{{\Omega _I}}}}}dx,
\end{align}
where $\vartheta _p^ *  = \max \left\{ {{\vartheta _1}, \ldots ,{\vartheta _p}} \right\}$ and ${\vartheta _p} = \frac{{{\gamma _{t{h_p}}}}}{{{\phi _p}\left( {{a_p} - {\gamma _{t{h_p}}}\sum\limits_{i = q + 1}^M {{a_i}} } \right)}}$.
\begin{proof}
We commence the diversity order analyses by characterizing the CDF $P_{ipSIC}^p$ in the high SNR regime. When $\rho  \to \infty $, the terms $\psi _p^ * $ and $\rho \psi _p^ * $ of $P_{ipSIC}^p$ are equal to zero and $\rho \vartheta _p^ * $, respectively. Upon substituting these terms into  \eqref{CD-NOMA:the OP of p-th user with ipSIC}, we can obtain \eqref{The asymptotic expression of the p-th user with ipSIC}. Noting that $P_{ipSIC}^{p,\infty }$ is a constant value with increasing the SNRs. The proof is completed.
\end{proof}
\end{corollary} 
\begin{remark}\label{remark2: the n-th user for CD-NOMA in case1 with ipSIC}
Upon substituting \eqref{The asymptotic expression of the p-th user with ipSIC} into \eqref{The definiation of diversity order}, the $p$-th terrestrial user with ipSIC achieves the zero diversity order. This is due to the influence of residual interference from ipSIC.
\end{remark}

\begin{corollary}\label{Corollary:the expression of asymptotic OP for the p-th user}
The asymptotic outage probability of the $p$-th terrestrial user with pSIC in the high SNR regime is given by
\begin{align}\label{The asymptotic expression of the p-th user}
P_{pSIC}^{p,\infty } = \frac{{M!}}{{\left( {M - p} \right)!p!}}\alpha _p^p{\left( {\psi _p^ * } \right)^p}\propto \frac{1}{{{\rho ^{p}}}},
\end{align}
where $ \propto $ denotes ``be proportional to". 
\begin{proof}
By invoking series representation \cite[Eq. (8.354.1)]{gradshteyn}, the term $\gamma \left( {k + 1,\psi _p^ * {\beta _p}} \right) $ of $P_{pSIC}^p$ can be further written as $\gamma \left( {k + 1,\psi _p^ * {\beta _p}} \right) = \sum\limits_{n = 0}^\infty  {\frac{{{{\left( { - 1} \right)}^n}{{\left( {\psi _p^ * {\beta _p}} \right)}^{k + 1 + n}}}}{{n!\left( {k + 1 + n} \right)}}} $. When $\rho  \to \infty $, that is $\psi _p^ *  \to 0$ and taking the first term $(n=0)$ of series representation, the asymptotic analysis of $\gamma \left( {k + 1,\psi _p^ * {\beta _p}} \right)$ is given by
\begin{align}\label{series representations}
\gamma \left( {k + 1,\psi _p^ * {\beta _p}} \right) \approx {\left. {\frac{{{{\left( {\psi _p^ * {\beta _p}} \right)}^{k + 1}}}}{{k + 1}}} \right|_{\psi _p^ *  \to 0}}.
\end{align} 

Upon substituting \eqref{series representations} into \eqref{CD-NOMA:the OP of p-th user}, the outage probability $P_{pSIC}^p$ can be approximated as
\begin{align}\label{series representations of process}
P_{pSIC}^p \approx & {\Theta _p}\sum\limits_{l = 0}^{M - p} {{
   {M - p}  \choose
   l  }} \frac{{{{\left( { - 1} \right)}^l}\alpha _p^{p + l}}}{{p + l}}\nonumber \\
 &  \times {\left( {\sum\limits_{k = 0}^\infty  {\frac{{{{\left( {{m_p}} \right)}_k}\delta _p^k{{\left( {\psi _p^ * } \right)}^{k + 1}}}}{{{{\left( {k!} \right)}^2}\left( {k + 1} \right)}}} } \right)^{p + l}}.
\end{align}
Based on \cite[Eq. (26)]{Yue8370069Unified} and further taking the first term of series representation in \eqref{series representations of process}, i.e., $k = 0$ and $l = 0$, we can obtain \eqref{The asymptotic expression of the p-th user}. The proof is completed.
\end{proof}
\end{corollary}
\begin{remark}\label{remark2: the n-th user for CD-NOMA in case1}
Upon substituting \eqref{The asymptotic expression of the p-th user} into \eqref{The definiation of diversity order}, the diversity order of $p$-th terrestrial user with pSIC is equal to $p$, which is closely related to the order of channel gains.
\end{remark}

\section{Numerical Results}\label{Numerical Results}
In this section, the numerical results are provided and show the impact of system parameters on NOMA-based satellite communication network. The links between satellite and terrestrial users are subject to Shadowed-Rician fading with channel parameters given in Table~\ref{parameter1} \cite{Abdi1198102}. Monte Carlo simulation parameters used in this section are summarized in Table~\ref{parameter} \cite{Bhatnagar6589295}. We assume that there are three users in the network, i.e., $M =3$. The power allocation factors for multiple users are set to be ${a_1} = 0.5$, ${a_2} = 0.4$, ${a_3} = 0.1$, respectively. Without loss of generality, the conventional OMA is selected to be a baseline, where the target rate ${R_o}$ of orthogonal user is equal to the sum rate of non-orthogonal users, ${R_1} = 0.1$, ${R_2} = 0.5$ and ${R_3} = 1$ bit per channel use.
\begin{table}[!t]
\centering
\caption{Table of Parameters for Satellite Communications Channel}
\tabcolsep6pt
\renewcommand\arraystretch{1.2} 
\begin{tabular}{|l|l|l|l|}
\hline
Shadowing  &  $b$  &  $m$  &  $\Omega$ \\
\hline
Frequent heavy shadowing (FHS)   &  0.063  &  0.739  &  $8.97 \times {10^{{\rm{ - }}4}}$ \\
\hline
Average shadowing (AS) &  0.126   &  10.1  &  0.835 \\
\hline
Infrequent light shadowing  (ILS)& 0.158  &   19.4 &  1.29 \\
\hline
\end{tabular}
\label{parameter1}
\end{table}
\begin{table}[!t]
\centering
\caption{Table of Parameters for Numerical Results}
\tabcolsep5pt
\renewcommand\arraystretch{1.1} 
\begin{tabular}{|l|l|}
\hline
Monte Carlo simulations repeated  &  ${10^5}$ iterations \\
\hline
Satellite orbit type  &   LEO \\  
\hline
Carrier frequency  &  $1 $ GHz  \\
\hline
3dB angle ${\varphi _{3{\rm{dB}}}}$ &   ${0.4^ \circ} $  \\
\hline
User's antenna gain per beam &   $3.5$ dBi  \\
\hline
Satellite's antenna gain per beam &   $24.3$ dBi  \\
\hline
The distance between satellite and users  &  $1000$ km \\
\hline
The angle between the beam center and users  &  ${0.1^ \circ }$ \\
\hline
\end{tabular}
\label{parameter}
\end{table}

\begin{figure}[t!]
    \begin{center}
        \includegraphics[width=2.9in, height=2.3in]{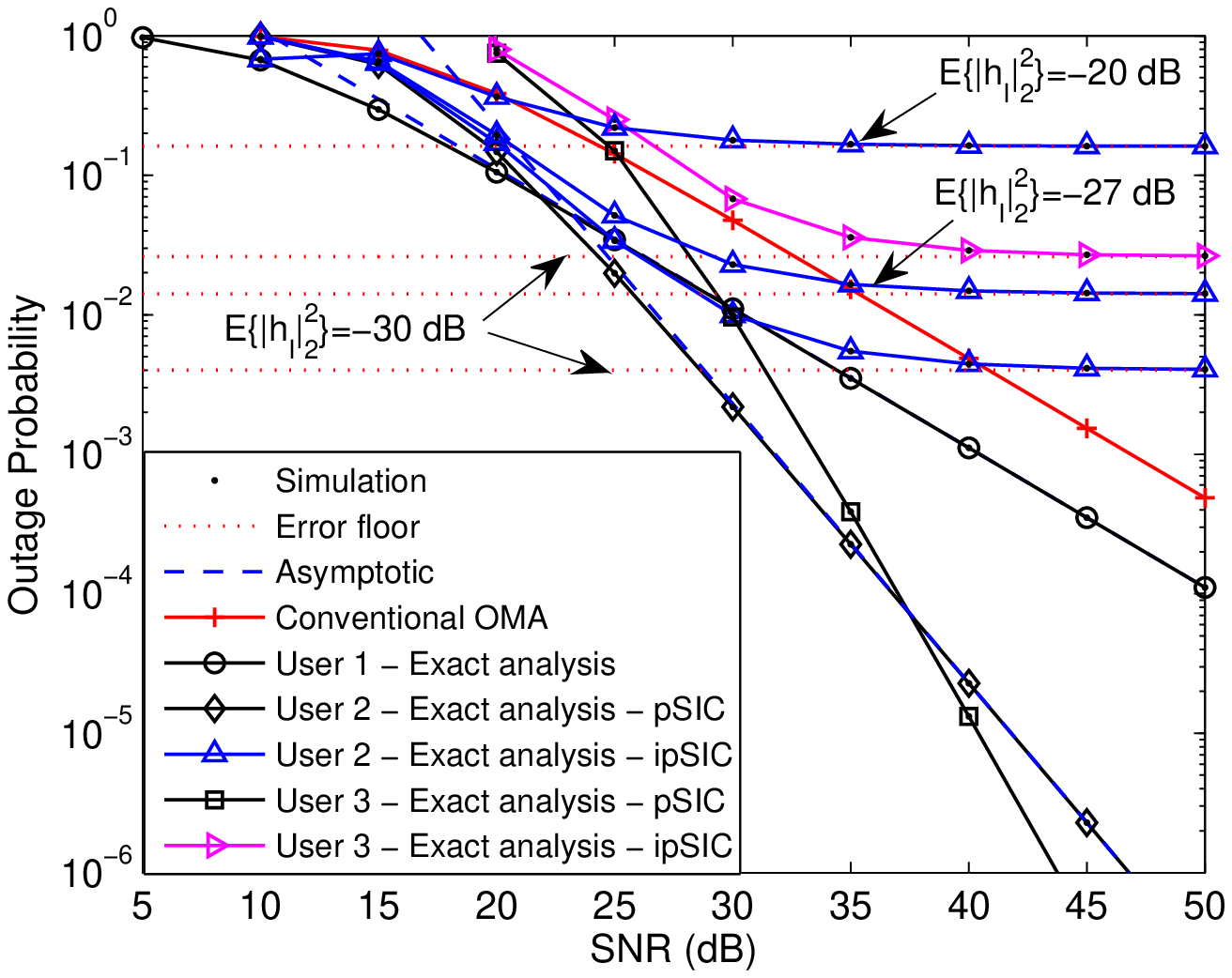}
        \caption{Outage probability versus the transmit SNR.}
        \label{The_Satellite_NOMA_OP_same_angle}
    \end{center}
\end{figure}
Fig. \ref{The_Satellite_NOMA_OP_same_angle} plots the outage probability versus the transmit SNR with satellite channel experiencing FHS. The exact outage probability of curves for the $p$-th terrestrial user (i.e., $p=1$, $p=2$ and $p=3$) with ipSIC/pSIC are given by numerical simulations and perfectly match with the analytical expressions. The asymptotic curves well approximate the exact outage probability curves. Due to the influence of residual interference, the outage probability of terrestrial users with ipSIC converge to an error floor. With increasing the value of residual interference, the outage behaviors of terrestrial user ($p=2$) with ipSIC are getting worse compared to other users. Another observation is that the outage performance of non-orthogonal users with pSIC is superior to that of orthogonal user. The basic reason for this phenomenon is that NOMA is capable of providing much more fairness when it serves multiple users at the same time \cite{Ding2014Randomly}.

\begin{figure}[t!]
    \begin{center}
        \includegraphics[width=2.9in, height=2.3in]{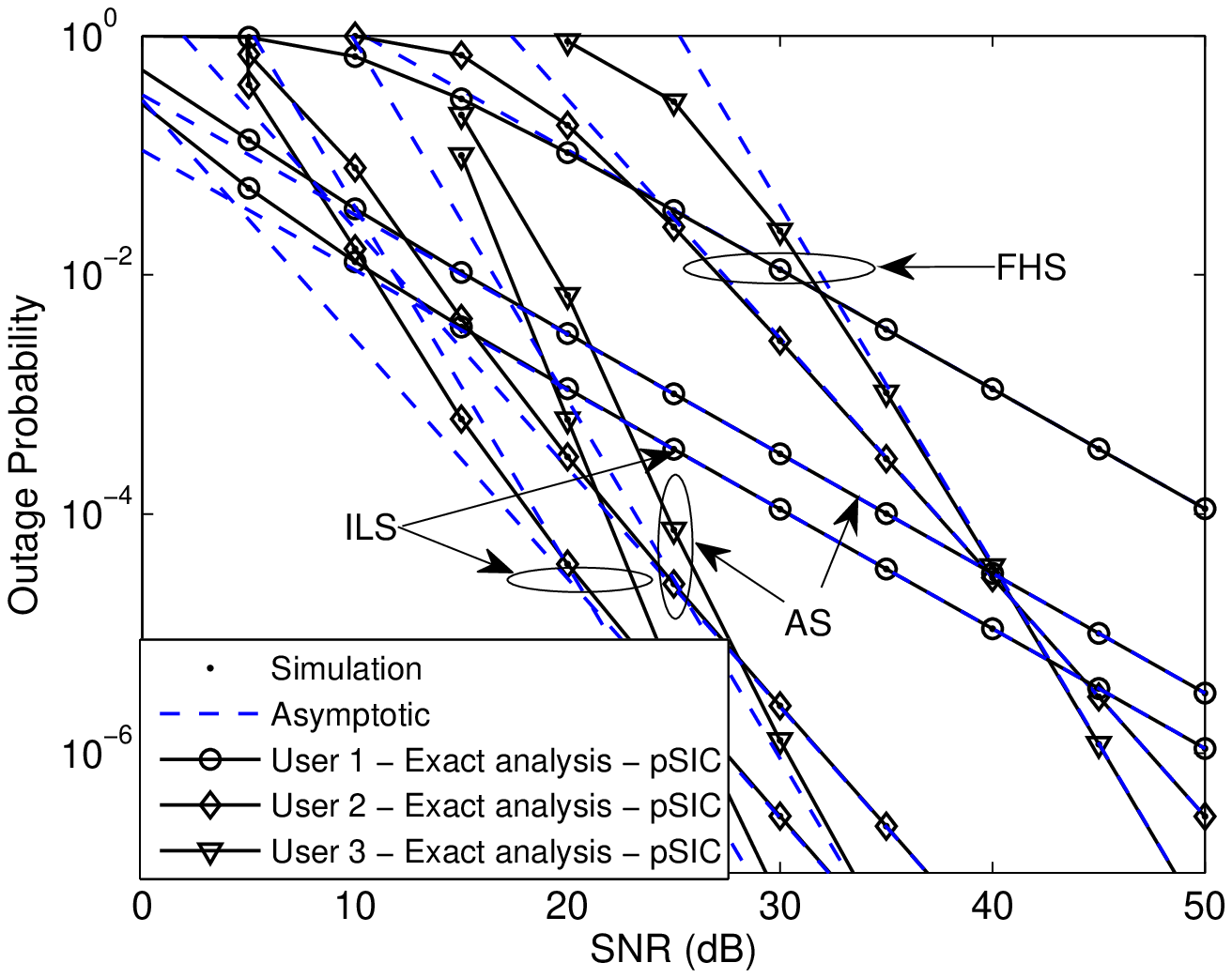}
        \caption{Outage probability versus the transmit SNR.}
        \label{Satellite_NOMA_OP_diff_channel_parameters}
    \end{center}
\end{figure}

Fig. \ref{Satellite_NOMA_OP_diff_channel_parameters} plots the outage probability versus the transmit SNR with different satellite channel parameters for the simulation setting ${\varphi _1} = {0.1^ \circ }$, ${\varphi _2} ={0.2^ \circ }$, ${\varphi _3} ={0.3^ \circ }$. We observe that the outage behaviors of users are sensitive to the shadowing condition of satellite-terrestrial channels. It is shown that the shadowing degrades network performance significantly. Frequent heavy shadowing results in a increasing outage performance, since the higher shadowing severities correspond to worse propagation conditions.
As the value of channel shadowing parameter, i.e., $b$, $m$, and $\Omega$ decreases, the outage performance of terrestrial users is becoming much worse seriously. This is due to the fact that both LoS component and multipath component become smaller for NOMA-based satellite network.

\begin{figure}[t!]
    \begin{center}
        \includegraphics[width=2.9in, height=2.3in]{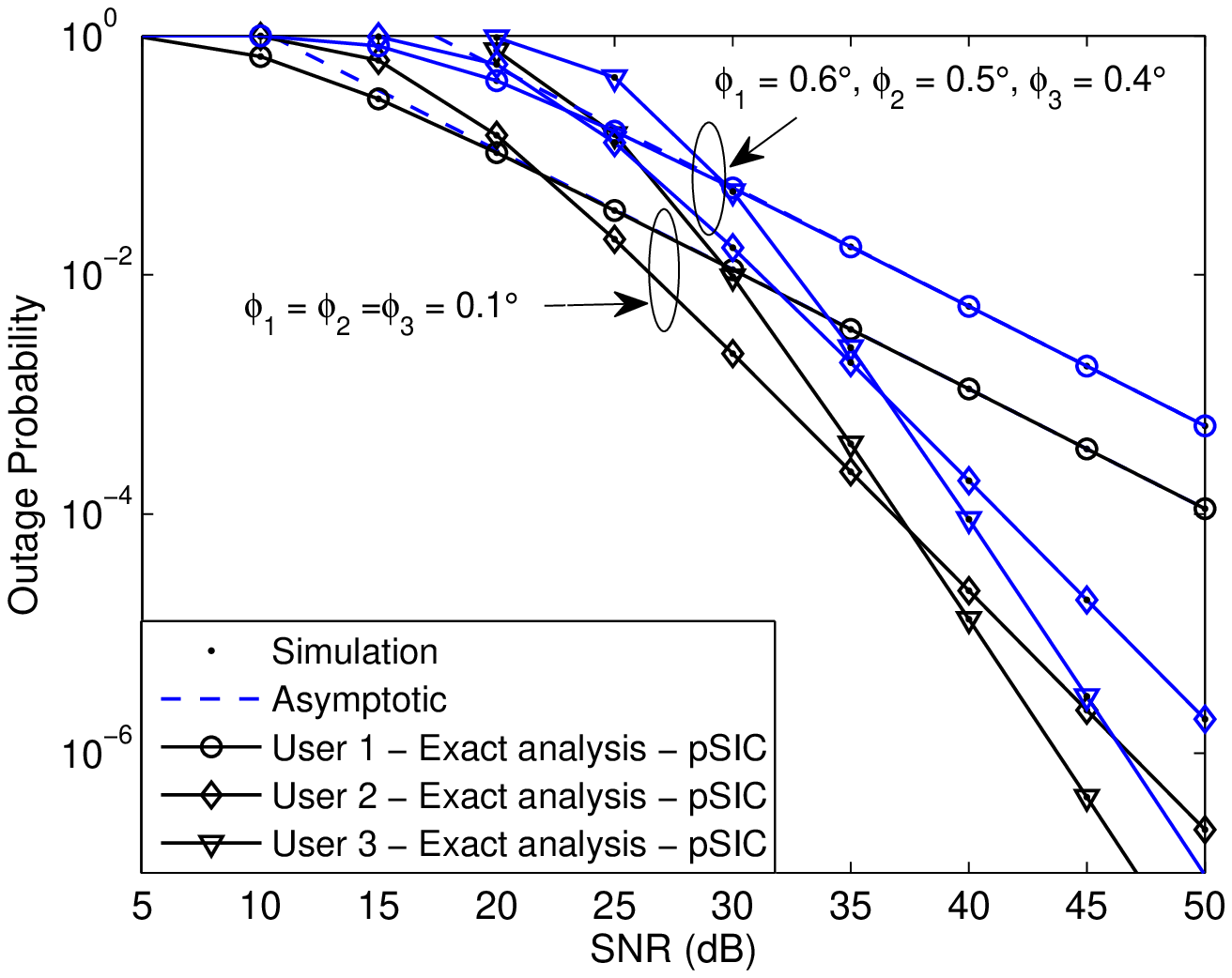}
        \caption{Outage probability versus the transmit SNR.}
        \label{Satellite_NOMA_OP_diff_angles}
    \end{center}
\end{figure}

Fig. \ref{Satellite_NOMA_OP_diff_angles} plots the outage probability versus the transmit SNR with different angles between the beam center and users for experiencing FHS. One can observe from figure that with the angles increasing, the outage behaviors of terrestrial users are becoming much worse. This is due to the fact that with the increase of angles, the users are getting closer to the edge of the beam relative to satellite. As a result, to obtain better system performance, we should adjust the angle of satellite to target the terrestrial
users from the perspective of service quality.
\section{Conclusion}\label{Conclusion}
In this paper, the application of NOMA to satellite communication network has been investigated over Shadowed-Rician Fading Channels.
The impact of system parameters on the performance of NOMA-based satellite network has been discussed, where the terrestrial users with ipSIC/pSIC are considered carefully.
New exact and asymptotic expressions of outage probability for terrestrial users have been derived to characterize the network performance. Simulation results have shown that the outage behaviors of NOMA-based satellite network with pSIC is superior to that of OMA.




\bibliographystyle{IEEEtran}
\bibliography{mybib}

\end{document}